\documentclass{sig-alternate}

\usepackage{amsmath,amssymb, amsfonts}
\usepackage{bm,cite,color,epsf,epsfig,graphicx,multirow,times,paralist,subfigure}
\usepackage{algorithm,algorithmic,bbm}

\newcommand{\ep}{\hfill $\Box$}

\newcommand{\el}{\end{flushleft}}
\newcommand{\bl}{\begin{flushleft}}

\newcommand{\separator}{
  \begin{center}
    \rule{\columnwidth}{0.3mm}
  \end{center}
}
\newenvironment{separation}{ \vspace{-0.3cm}  \separator  \vspace{-0.2cm}}
{  \vspace{-0.4cm}  \separator  \vspace{-0.1cm}}

\newtheorem{theorem}{Theorem}[section]
\newtheorem{lemma}[theorem]{Lemma}
\newtheorem{corollary}[theorem]{Corollary}

\def\pc{\rm{pc}}
\def\sched{\rm{sched}}
\def\*{{\star}} 

\DeclareGraphicsExtensions{.pdf,.eps}

\graphicspath{{.//}}

\begin{document}

%\title{Iterative Power Packing for Distributed Multiple Access in Wireless Networks}

\title{Optimal Distributed Scheduling in Wireless Networks under SINR Interference Model}

\author{P. Chaporkar${}^\star$, A. Proutiere${}^\dagger$\thanks{${}^\star$ IIT Mumbai; ${}^\dagger$KTH, The Royal Institute of Technology.}}

\maketitle

\begin{abstract}
Radio resource sharing mechanisms are key to ensuring good performance in wireless networks. In their seminal paper \cite{tassiulas1}, Tassiulas and Ephremides introduced the Maximum Weighted Scheduling algorithm, and proved its throughput-optimality. Since then, there have been extensive research efforts to devise distributed implementations of this algorithm. Recently, distributed adaptive CSMA scheduling schemes \cite{jiang08} have been proposed and shown to be optimal, without the need of message passing among transmitters. However their analysis relies on the assumption that interference can be accurately modelled by a simple interference graph. In this paper, we consider the more realistic and challenging SINR interference model. We present {\it the first distributed scheduling algorithms that (i) are optimal under the SINR interference model, and (ii) that do not require any message passing}. They are based on a combination of a simple and efficient power allocation strategy referred to as {\it Power Packing} and randomization techniques. We first devise algorithms that are rate-optimal in the sense that they perform as well as the best centralized scheduling schemes in scenarios where each transmitter is aware of the rate at which it should send packets to the corresponding receiver. We then extend these algorithms so that they reach throughput-optimality.
\end{abstract}

\section{Introduction}

The throughput experienced on a given link in wireless networks is affected by the interference generated by the transmitters of other links. Interference management constitutes the main issue in the design of simple and efficient resource allocation (or Multiple Access Control) algorithms for such networks. Solving this issue becomes even more challenging when links have to share radio resources in a distributed manner. Distributed power control \cite{zander, foschini} is often used (e.g. in cellular systems) to tackle this issue. However, when links strongly interfere each other, power control is inefficient as the set of rates that can be simultaneously achieved on the competing links exhibits non-convexities. For such scenarios, scheduling transmissions over time is much more efficient and results in a much larger rate region. Most existing MAC algorithms for WLANs, Mesh, and AdHoc networks are {\it scheduling} algorithms: transmitters only decide when to be active, and when active, they use a single power level, often the maximum power level. In their seminal paper \cite{tassiulas1}, Tassiulas and Ephremides proposed the queue-length based {\it Maximum Weighted Scheduling} (MWS) algorithm, and proved its throughput-optimality (meaning that it can stabilize the network whenever this is at all possible). However the MWS algorithm is centralized, and often requires to repeatedly solve instances of NP-hard optimization problems. 

Over the last two decades, there have been important research efforts towards the design of low-complexity and distributed versions of the MWS algorithm (refer to the related work section for references). Recently, in \cite{jiang08,shah,walrand,q-csma}, simple and throughput-optimal adaptive versions of CSMA have been proposed. These algorithms enjoy the property of being fully distributed, in the sense that they do not require any kind of message passing among the various transmitters. However their analysis and performance guarantees rely on the strong assumption that interference can be modelled as a simple undirected graph (in the interference graph, vertices represent links, and an edge between two links mean that these links cannot be simultaneously activated). In particular, this simplistic interference model cannot account for the well-known hidden and exposed terminal problems, and more generally does not accurately capture the very nature of interference. In this paper, we revisit the design of efficient and distributed MAC protocols under the more realistic SINR interference model. Specifically, we aim at answering the following question:

{\it Can we devise fully distributed and optimal scheduling algorithms for wireless networks under the SINR interference model?}

By {\it fully distributed}, we mean that transmitters are not allowed to exchange any signalling message, and the only feedback available at a given transmitter is the level of interference measured at the corresponding receiver (just as in classical distributed power control mechanisms \cite{zander, foschini}). {\it Optimal} may have several meanings. To discuss the different versions of optimality, let us first introduce the notion of rate region defined as the set of rates that can be simultaneously achieved on the various links using some (centralized) scheduling algorithms. (i) Rate-optimality: in this case, transmitters always have packets to send, i.e., they are fully backlogged. An algorithm is rate-optimal, if it can achieve any rate vector within the rate region. (ii) Throughput-optimality: in this case, each transmitter receives, in its (infinite) buffer, packets arriving according to a stationary ergodic process with fixed average rate. An algorithm is throughput-optimal if it stabilizes\footnote{The assumptions made on the packet arrival processes and the notion of stability are described in Section 7.} all buffers as long as the mean arrival rate vector belongs to the largest open set contained in the rate region. 

In this paper, we show that surprisingly, the answer to the above question is positive, and develop fully distributed and rate-optimal scheduling algorithms. We also demonstrate how these algorithms can be used towards the design of throughput-optimal scheduling schemes. In the proposed framework, we first divide time into frames consisting of a fixed number of slots. Each transmitter is then allowed to adapt the power levels used in the various slots of a frame to achieve the rate it is targeting. Our solution is based on a simple power control mechanism, referred to as {\it Power Packing} (PP). Under this mechanism, each transmitter aims at achieving its target rate while minimizing the number of slots actually used, hence leaving as many radio resources as possible to the other transmitters. PP algorithms are shown to be rate-optimal when two links compete for the use of resources. However, in more general networks and in some rare scenarios, they may fail at achieving certain rate vectors that could have been realized using centralized scheduling. By just adding to the algorithms some level of randomization in the power allocation, we overcome this issue and recover rate-optimality. All the proposed algorithms are simple and do not require any message passing: each transmitter adapts its power levels in the various slots depending on the observed interference levels. To our knowledge, the proposed algorithms constitute the first scheduling schemes that are fully distributed (no message passing) and optimal under the SINR interference model.   

The paper is organized as follows:\\
(i) In Section 2, we present a brief overview of the existing literature on distributed resource allocation algorithms in wireless networks. \\
(ii) In Sections 3 and 4, we present our generic framework, Power Packing algorithms and explain their rationale.\\
(iii) We establish the rate-optimality of Iterative Power Packing algorithms for 2-link networks in Section 5.\\
(iv) For more general networks, we explain, in Section 6, why Iterative Power Packing algorithms may in some rare cases fail. To solve this issue, we introduce some Perturbed versions of Iterative PP algorithms and show their rate-optimality.\\
(v) In Section 7, we show how our rate-optimal algorithms can be adapted to achieve throughput-optimality.\\
vi) Finally, in Section 8, we illustrate the efficiency of our algorithms using numerical experiments.

\section{Related work} 

There have been, over the last two decades, a tremendous research effort towards the design of distributed resource sharing mechanisms in wireless networks under various interference models (see e.g. surveys \cite{tassiulassurvey,jiangsurvey}). For the simplistic interference graph model, researchers have developed scheduling algorithms that implement the celebrated throughput-optimal MWS algorithm \cite{tassiulas1} in a distributed manner. Some of these algorithms use message passing, see e.g. \cite{modiano2}, some others do not require message passing, e.g. as the adaptive versions of CSMA, see e.g. \cite{jiang08,shah,walrand,q-csma}. 

In this paper, we are interested in the more realistic SINR interference model. This model has also attracted a lot of attention recently, see e.g. \cite{neely,evans,yeh,modiano}. For example in \cite{evans}, the authors derive utility-optimal power control schemes, but the achieved rate region is restricted to that achieved by power control only. In \cite{yeh, modiano}, the authors design schemes also enabling time sharing, and hence scheduling. These schemes implement the MWS algorithm, but require message passing (basically, a transmitter need to know its impact on the throughputs on other links). In a series of papers \cite{bambos1,bambos2,bambos3}, Bambos et al. design power control algorithms that ressemble Foschini-Miljanic algorithm \cite{zander, foschini} in the sense that the power update at a transmitter only depends on the measured interference level, and on some local queue size. These schemes are fully distributed, and seem to realize time sharing when needed. However, their optimality has not been established, and there may be network examples where these schemes are not optimal. 

\section{Models and Preliminaries}

\subsection{Network model}

We consider a network consisting of $N$ interfering links (transmitter-receiver pairs). We are primarily interested in the design of rate-optimal algorithms, and so each link $i$ has a target rate requirement $R_i^t$ (corresponding to the QoS requirements of the underlying application). To achieve this target rate, link-$i$ transmitter may adapt its transmission power $p_i$. The transmission power at any transmitter cannot exceed $P_{\max}$. Links interfere, and we assume here that each receiver treats interference as noise. Let $g_{ji}$ denote the channel gain from link-$j$ transmitter to link-$i$ receiver. Thermal noise is Gaussian, with power $N_0$. Under these assumptions, the maximum rate that link $i$ can achieve can be written as: $r_i(p)=f\left({g_{ii}p_i\over N_0+\sum_{j\neq i}g_{ji}p_j}\right)$, where $f(\cdot)$ is an increasing positive concave function, typically $f(x)=W\log(1+x)$, and $p=(p_1,\ldots,p_N)$.

\medskip
\noindent
{\bf Notation.} Let ${\cal U}$ be a subset of $\mathbb{R}_+^N$. We denote by $\rm{conv}({\cal U})$ the convex hull of ${\cal U}$, and by $\partial{\cal U}$ the Pareto-boundary of ${\cal U}$: $x\in \partial{\cal U}$ iff $x\in {\cal U}$ and $\forall y\in {\cal U}$, $y\ge x$ coordinate-wise implies that $x=y$. We further define $\bar{\cal U}=\{r\in \mathbb{R}_+^N: \exists R\in {\cal U}, \forall i, r_i\le R_i\}$ as the smallest coordinate-convex set containing ${\cal U}$. ${\bf 1}=(1,\ldots,1)$.

\subsection{Power control vs. Scheduling}

\begin{figure}[htb]
\begin{center}
\scalebox{0.5}{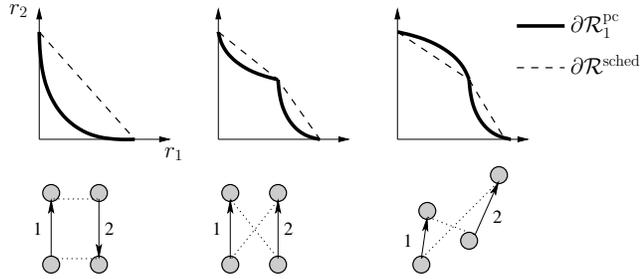}
\end{center}
\caption{Rate regions under power control and scheduling: strong (left) and weak (middle and right) interference cases.}
\label{fig:rr1}
\end{figure}

We define ${\cal R}_1^{\pc}=\{ r(p): \forall i, p_i\in [0,P_{\max}]\}$ as the set of vectors representing rates that can be achieved on the various links using power control. This set is known to be non-convex, and may exhibit different types of shapes, depending on the values of gains $(g_{ij},i,j)$, i.e., on the network geometry. Let ${\cal S}_1=\{ r(p): \forall i, p_i\in \{ 0,P_{\max}\}\}$ be the set of vectors representing link rates achieved using {\it binary} power control, i.e., for any $i$, link-$i$ transmitter either remains silent or transmits at maximum power $P_{\max}$. ${\cal S}_1$ is referred to as the set of {\it schedules}. The set ${\cal R}^{\sched}$ of link rates that can be achieved by switching schedules over time is the convex hull of ${\cal S}_1$: ${\cal R}^{\sched}=\rm{conv}({\cal S}_1)$. Now we may allow transmitters to use both power control and time sharing. In this case, the set of achievable rate vectors is ${\cal R}=\rm{conv}({{\cal R}_1^{\pc}})$. In general, both power control and time sharing are required, in the sense that we may have for the same network: ${\cal R}_1^{\pc}\subsetneq {\cal R}$ and  ${\cal R}^{\sched}\subsetneq {\cal R}$. We illustrate these observations in Figure \ref{fig:rr1}, where we depict the Pareto-boundaries $\partial{\cal R}_1^{\pc}$ and $\partial{\cal R}^{\sched}$ of the set ${\cal R}_1^{\pc}$ and ${\cal R}^{\sched}$, respectively, for different interference scenarios. When links strongly interfere each other, time sharing (scheduling) is enough, whereas when interference becomes weaker, power control may be necessary. In this paper, our goal is to design fully distributed algorithms enabling the various links to reach their target rates $R^t=(R_1^t,\ldots,R_N^t)$, provided that $R^t\in {\cal R}^{\sched}$.   

\subsection{Multi-slot systems}

To share radio resources among links, we divide time into frames. Each frames consists of a fixed number $M$ of time slots of equal durations. If each transmitter is allowed to use different power levels on the various slots, the rates $R(p)$ achieved on the various links can be written as:
$$
R_i(p)={1\over M}\sum_{m=1}^Mf({g_{ii}p_{im}\over N_0+\sum_{j\neq i}g_{ji}p_{jm}}),\quad \forall i,
$$
where $p=(p_{im},i=1,\ldots,N, m=1,\ldots,M)$ and $p_{im}$ is the power level used by link-$i$ transmitter on the $m$-th slot in each frame. The set of achievable rates using such multi-slot power control is then: ${\cal R}_M^{\pc} =\{ R(p): \forall i, \forall m, p_{im}\in [0,P_{\max}]\}$. ${\cal R}_M^{\pc}$ can also be expressed as combinations of rate vectors in ${\cal R}_1^{\pc}$: ${\cal R}_M^{\pc} =\{ r: \exists s_m\in {\cal R}_1^{\pc}, m=1,\ldots,M: r={1\over M}\sum_{m=1}^Ms_m \}$. Observe that we do not impose any constraint on the total power used by a transmitter per frame. 

\begin{figure}[htb]
\begin{center}
%\scalebox{0.28}{\input{rater3.pdf_t}}
\includegraphics[width=8cm]{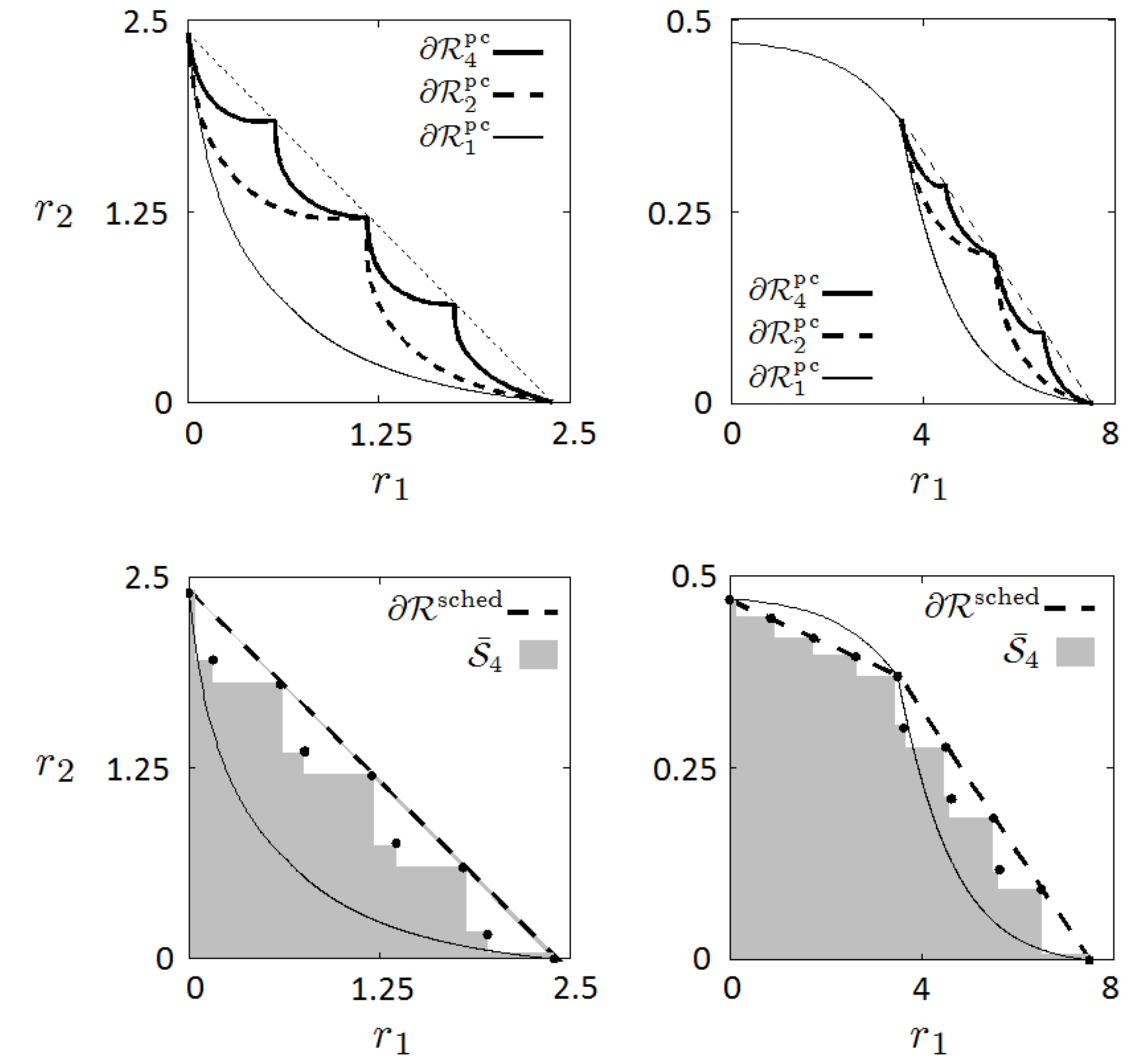}
\end{center}
\caption{Rate regions under power control (top-right, top-left) and scheduling (bottom-right, bottom-left) in multi-slot systems. For scheduling, the black dots correspond to the Pareto-boundary of ${\cal S}_4$. The network for figures on the left (resp. right) corresponds to gains $g_{ij}=1$, $\forall i,j$ (resp. $g_{11}=2000$, $g_{12}=0.4$, $g_{22}=0.6$, $g_{21}=0.4$), $N_0=0.1$, $P_{\max}=1$.}
\label{fig:rr2}
\end{figure}

Now consider scenarios where transmitters are allowed, in a given slot, either to use maximum power $P_{\max}$ or to remain silent. As earlier, we may define a set ${\cal S}_M$ of {\it schedules}:
$$
{\cal S}_M=\{ R(p): \forall i,\forall m, p_{i,m}\in \{ 0,P_{\max}\}\}.
$$ 
Sharing time among the various schedules in ${\cal S}_M$ increases the set of achievable rates, i.e., ${\cal S}_1\subset {\cal S}_M$. Observe that the set of achievable rates on the various links using a single schedule in ${\cal S}_M$ is $\bar{\cal S}_M$ is the smallest coordinate convex set containing ${\cal S}_M$. The various notions of rate regions and their Pareto-boundaries are illustrated in Figure \ref{fig:rr2}.

Note that designing distributed radio resource allocation schemes achieving any $R^t\in {\cal R}^{\sched}$ is difficult for this requires to identify the various proportions of time schedules in ${\cal S}_M$ are used. Designing schemes achieving any $R^t\in \bar{\cal S}_M$ may seem easier because in this case we only need to identify a single schedule in ${\cal S}_M$ satisfying the rate requirements. 

As stated in the following lemma, when the number of slots per frame is large, we can achieve the largest rate region ${\cal R}$ by just implementing power control per slot, and ${\cal R}^{\sched}$ by choosing a fixed schedule from ${\cal S}_M$. All proofs are presented in appendix.

\begin{lemma}\label{lem:rr} $\lim_{M\to\infty}{\cal R}_M^{\pc}={\cal R}$,  $\lim_{M\to\infty}\bar{\cal S}_M={\cal R}^{\sched}$. 
\end{lemma}

Here $\lim_{M\to\infty}A_M=B$ means that for every point $R\in B$, there exists a sequence of points $(X_M,M\ge 1)$ such that $X_M\in A_M$ for all $M$, and $\lim_{M\to\infty}X_M=R$.

In practice, we observe that the introduction of frames, even of small sizes, considerably increases the rate region: in other words, the sequence of sets $\bar{\cal S}_M$, $M=1,2,...$ rapidly approaches ${\cal R}^{\sched}$. Based on this observation and on previous lemma, we use the following strategy to design distributed resource allocation schemes approximately achieving rates in ${\cal R}^{\sched}$: (i) We select a frame size $M$ so that $\bar{\cal S}_M$ provides a good approximation of ${\cal R}^{\sched}$, e.g. $M=16$; (ii) we devise distributed resource allocation schemes achieving any rate vector in $\bar{\cal S}_M$. 

\section{Power Packing}

In this section, we present power packing algorithms for the multi-slot systems introduced in the previous section. When executing such algorithm, a transmitter aims at minimizing the number of slots actually used (a slot is {\it used} on a link, if the corresponding transmitter selects a strictly positive power level in this slot) while achieving the target rate. To run power packing algorithms, transmitters just need to measure the interference generated by other transmitters in the slots composing a frame. 

\subsection{Algorithms}

Let $I_{im}(p)$ denote the interference perceived at link-$i$ receiver during the $m$-th slot of the frame, given the power allocation $p=(p_{jm})_{j,m}$: $I_{im}(p)=N_0+\sum_{j\neq i}g_{ji}p_{jm}$. We also introduce $h_i:[0,P_{max}]^M\times \mathbb{R}_+^M\to \mathbb{R}_+$ that gives the rate on link $i$ as a function of link-$i$ transmitter power levels, and perceived interference levels in the various slots: $h_i(p_i,I_i)={1\over M}\sum_{m=1}^M f\left( {p_{im}g_{ii}\over I_{im}}\right)$. 

\subsubsection{Power Packing (PP) algorithm}
Power packing algorithm is executed by a transmitter in response to the observed interference levels in the various slots of a frame. The principle of power packing is to sequentially fill with power slots in increasing order of perceived interference and until the target rate is reached. If the latter cannot be reached, the transmitter just remains silent in all slots. The algorithm, whose pseudo-code is presented below, is illustrated in Figure \ref{fig:pp}. 

\begin{figure}[htb]
\begin{center}
\includegraphics[width=8cm]{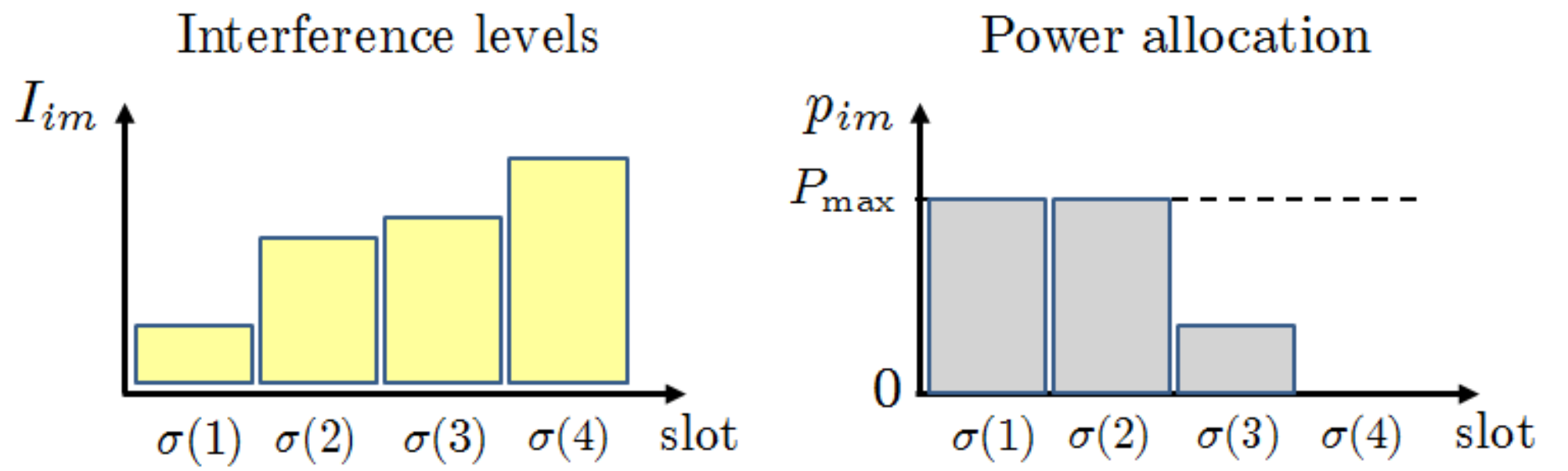}
\end{center}
\caption{Example of power allocation obtained after power packing algorithm - frame size $M=4$. }
\label{fig:pp}
\end{figure}

\begin{separation}
\vspace{-0.07cm}
{\bf PP algorithm.}  (Executed at link-$i$ transmitter)
%\vspace{-0.5cm}\separator
%\vspace{-0.2cm}
\begin{itemize}
\item[Input:] target rate $R_i^t$, interference levels $I_i=(I_{im})_m$.
\item[1.] Compute the rate $\bar{R}_i=h_i(P_{\max}{\bf 1},I_i)$ achieved using maximum power in each slot,
\item[2.] If $\bar{R}_i< R_i^t$: select power allocation $p_i=(0,\ldots,0)$;
\item[3.] If $\bar{R}_i\ge R_i^t$: order slots in increasing interference levels: let $\sigma$ be a permutation of $\{1,\ldots,M\}$ such that $I_{i\sigma(1)}\le \ldots,I_{i\sigma(M)}$. Define 
$$
\tilde{m} = \min\{m: {1\over M}\sum_{k=1}^m f({P_{\max}g_{ii}\over I_{i\sigma(k)}})\ge R_i^t\}.$$ 
Select the unique power allocation $p_i$ such that: $\forall m< \tilde{m}$,  $p_{i\sigma(m)}=P_{\max}$, $\forall m > \tilde{m}, p_{i\sigma(m)}=0$, and $R_i^t=h_i(p_i,I_i)$.
\end{itemize} 
\end{separation}

\subsubsection{Binary Power Packing (BPP) algorithm}
The PP algorithm has a {\it binary} version, where the transmitter is allowed in a given slot to either use full power $P_{\max}$ or remain silent. BPP algorithm is identical to PP algorithm, except for step 3 where the power allocation differs: If $\bar{R}_i\ge R_i^t$, the transmitter uses the power allocation $p_{i\sigma(m)}=P_{\max}1_{\{ m\le \tilde{m}\}}$.

\subsection{Game theoretical interpretation} 

We now provide a game theoretical perspective on PP and BPP algorithms. Consider a noncooperative game played by the $N$ transmitters. Each transmitter competes rationally against the others by selecting a power allocation across the $M$ available slots. The set of strategies available to any transmitter consists of all possible power allocation across slots. In the case where transmitters can use any power level between 0 and $P_{\max}$, the set of strategies is ${\cal P}=\{ p_i: \forall m, p_{im}\in [0,P_{\max}] \}$, whereas in case of binary power control, this set reduces to ${\cal P}_B=\{ p_i: \forall m, p_{im}\in \{ 0,P_{\max} \}\}$. The utility function $U_i(p)$ of transmitter $i$ is defined as follows: $U_i(p)=C\times 1_{\{ R_i(p)\ge R_i^t\} } - \sum_{m=1}^M {p_{im}g_{ii}\over I_{im}(p)},$ where $C$ is a positive constant such that $C>\sum_{m=1}^MP_{\max}g_{ii}/N_0$, for any link $i$. We denote by ${\cal G}(R^t)$ (resp. ${\cal G}_B(R^t)$) the game defined above when the set of strategies is ${\cal P}$ (resp. ${\cal P}_B$). It can be easily shown that with our choice of utility functions, the PP and the BPP algorithms executed by link-$i$ transmitter can be interpreted as the best response to the power allocations $p_{-i}=(p_{jm})_{j\neq i,m}$ used by the other transmitters. In other words, assume that the power allocation $p_j$ used by link-$j$ transmitter is fixed for all $j\neq i$. These allocations result in interference levels $(I_{im})_m$ at link-$i$ receiver. For example, the power allocation obtained when link-$i$ transmitter executes PP algorithm under these conditions solves the following optimization problem: maximize $U_i(q_i,p_{-i})$, over $q_i\in {\cal P}$.

\section{Two link case: Iterative Power Packing}

In this section, we restrict our attention to two-link networks. We propose and analyze the convergence of Iterative Power Packing (IPP) algorithms. The latter consist in letting transmitters sequentially update their power allocation using PP or BPP algorithms. 

\subsection{IPP and IBPP algorithms}

To define IPP and IBPP algorithms, we first introduce a sequence $s=(s[t])_{t\ge 1}$, $s[t]\in \{1,2\}$, defining the order in which transmitters update their power allocation. We assume that the sequence satisfies the following property, stating that each transmitter gets to update its power allocation an infinite number of times:
\begin{itemize}
\item[(P1)] $\forall t\ge 1$, $\exists t_1,t_2 \ge t$: $s[t_1]=1$ and $s[t_2]=2$.
\end{itemize}
This property is referred to as {\it liveness} property in game theory. A sequence of updates satisfying this property is in principle easy to generate in a distributed manner, for example using independent Poisson clocks with identical rate at the various transmitters. Refer to \textsection 6.1 for more details. We are now ready to define IPP algorithm:

\begin{separation}
\vspace{-0.07cm}
{\bf IPP algorithm.} 
%\vspace{-0.5cm}\separator
%\vspace{-0.2cm}
\begin{itemize}
\item[Input:] target rate vector $R^t$, update sequence $s$, initial power allocation $p[0]$.
\item[For each step $t\ge 1$:] Let $i=s[t]$.
\item[1.] Link-$i$ transmitter measures interference levels $I_{i}(p[t-1])=(I_{im}(p[t-1])_m$ in the different slots;
\item[2.] Link-$i$ transmitter runs PP algorithms with inputs $R_{i}^t$ and $I_{i}(p[t-1])$.
\end{itemize} 
\end{separation}

IPP algorithm has a binary version, IBPP algorithm, obtained by just replacing PP algorithm by the BPP algorithm in the above pseudo-code. IPP and IBPP algorithms correspond to the best response dynamics or Nash dynamics of the games ${\cal G}(R^t)$ and ${\cal G}_B(R^t)$, respectively. They can easily be implemented in a fully distributed manner: when a transmitter updates its power allocation, it only needs to measure interference levels on the various slots and to know its own target rate. 

\subsection{Convergence}

To study the convergence of IPP and IBPP algorithms, we introduce the notion of {\it repulsive} power allocation. We say that $p=(p_1,p_2)\in [0,P_{\max}]^{2M}$ is repulsive if and only if there exist a permutation $\sigma$ of $\{1,\ldots,M\}$ and two integers $m_1,m_2\in\{0,1,\ldots,M+1\}$ such that for all $m\in \{1,\ldots,M\}$ (i) $m \le m_1$ implies $p_{1\sigma(m)}=P_{\max}$, and $m>m_1+1$ implies $p_{1\sigma(m)}=0$, (ii) $m \ge m_2$ implies $p_{2\sigma(m)}=P_{\max}$ and $m< m_2-1$ implies $p_{2\sigma(m)}=0$.

The set of rate vectors that can be achieved using repulsive power allocation is then defined as:
$$
{\cal R}_M^{\rm{IPP}}=\{ R\in \mathbb{R}_+^2: \exists p\hbox{ repulsive: }R=R(p) \}
$$
In the case the power allocation is binary, we similarly define:
$$
{\cal R}_M^{\rm{IBPP}}=\{ R\in \mathbb{R}_+^2: \exists p\in \{0,P_{\max}\}^{2M}\hbox{ repulsive: }R\le R(p) \}
$$
In what follows, we show that ${\cal R}_M^{\rm{IPP}}$ (resp. ${\cal R}_M^{\rm{IBPP}}$) is the rate region achieved under IPP (resp. IBPP) algorithm.

\medskip
\begin{theorem}\label{th:ipp}
Let $R^t\in {\cal R}_M^{\rm{IPP}}$ (resp. $\in {\cal R}_M^{\rm{IBPP}}$). Then from any initial power allocation, IPP (resp. IBPP) algorithm converges to a repulsive power allocation $p\in [0,P_{\max}]^{2M}$ (resp. $p\in \{0,P_{\max}\}^{2M}$) such that $R^t=R(p)$ (resp. $R^t\le R(p)$).
\end{theorem}

\begin{figure}[htb]
\begin{center}
\includegraphics[width=8cm]{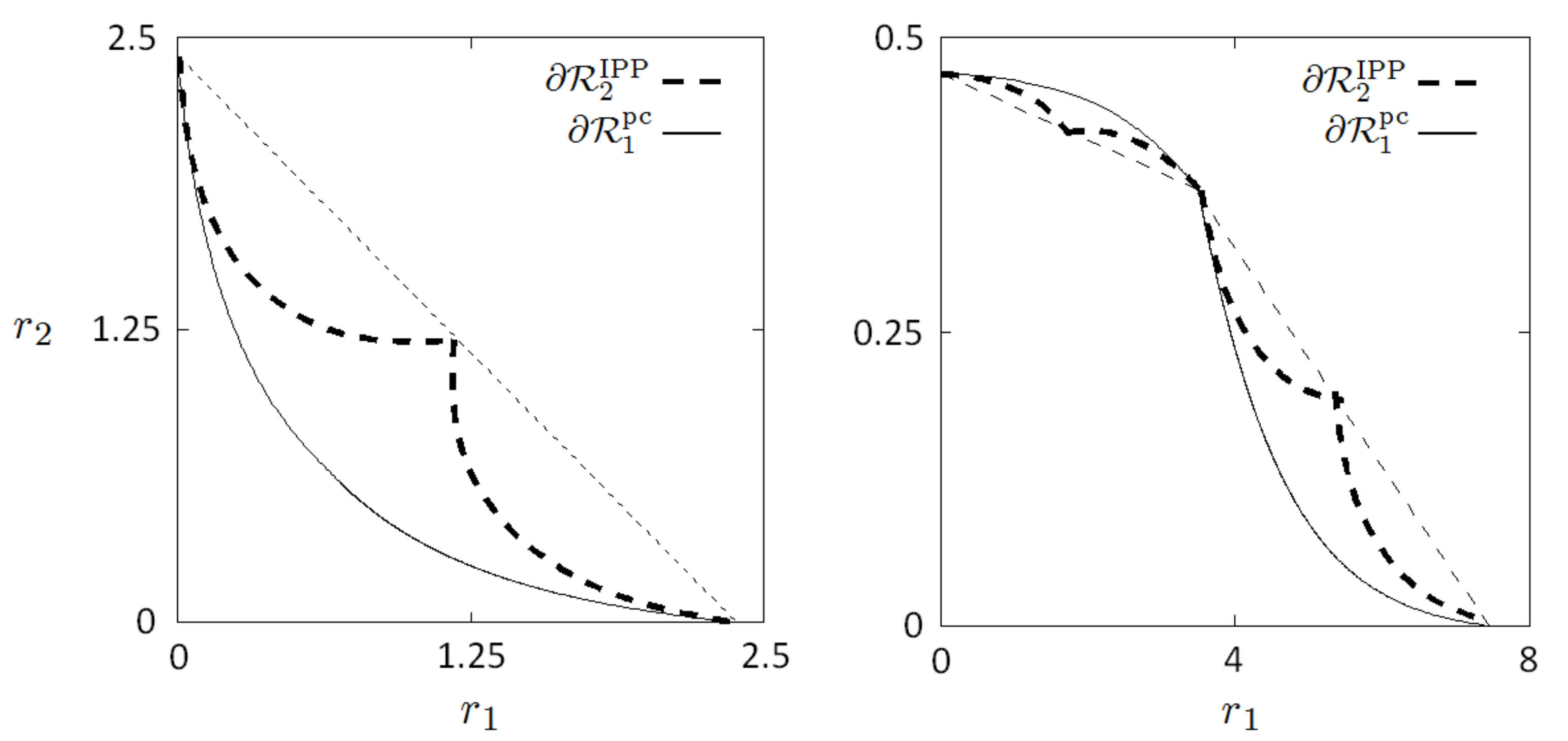}
\end{center}
\caption{Rate regions obtained through IPP algorithm. }
\label{fig:ipp}
\end{figure}

\medskip 
From a game theoretical perspective, this result states that if $R^t\in {\cal R}_M^{\rm{IPP}}$, then the Nash dynamics converge to a pure Nash Equilibrium corresponding to a repulsive power allocation and achieving the target rates $R^t$. Figure \ref{fig:ipp} illustrates the rate regions ${\cal R}_M^{\rm{IPP}}$. 

We do not represent ${\cal R}_M^{\rm{IBPP}}$, because in fact, ${\cal R}_M^{\rm{IBPP}}=\bar{\cal S}_M$. This can be shown by applying the following argument: Let $p$ be a binary power allocation; modify this allocation such that (i) the number of slots used by each transmitter is not changed, and (ii) the new allocation is repulsive. It is easy to see that the new allocation provides greater rates to all transmitters. notice however that ${\cal R}_M^{\rm{IPP}}$ is smaller than ${\cal R}_M^{\rm{pc}}$: this is true when the initial rate region ${\cal R}_1^{\rm{pc}}$ has {\it concave} parts -- see Fig. \ref{fig:ipp} (right). In this case, some points of the Pareto-boundary of ${\cal R}_M^{\rm{pc}}$ can only be achieved by non-repulsive power allocations of the type $p=(x{\bf 1},P_{\max}{\bf 1})$, where $x\in [0,P_{\max}]$.

To conclude this section, one can show (as in Lemma 1) that ${\cal R}_M^{\rm{IPP}}$ and ${\cal R}_M^{\rm{IBPP}}$ approximate ${\cal R}^{\sched}$ (when $M$ is large). 

\begin{lemma} $\lim_{M\to\infty} {\cal R}_M^{\rm{IPP}}={\cal R}^{\sched}$, $\lim_{M\to\infty}{\cal R}_M^{\rm{IBPP}}=\bar{\cal R}^{\sched}$.
\end{lemma}

As a consequence, any target rate vector $R^t$ inside ${\cal R}^{\sched}$ can be achieved either using IPP or IBPP algorithm, provided that the frame size is large enough. In other words, IPP or IBPP algorithms are approximately rate-optimal in 2-link networks (we will give a more precise definition of what we mean by "approximately rate=optimal" in the next section).

\section{Multiple link case: Iterative Perturbed Power Packing}

In this section, we consider general networks with more than two links. We first explain why IPP or Binary-IPP may fail at converging for some specific target rates in ${\cal R}^{\sched}$. We then present two binary power control algorithms to overcome this issue. 

An example of networks and target rates where IPP does not work is as follows. Consider a network consisting of 3 links sharing the same receiver (Access Point scenario), and let $M=3$ slots. The two first transmitters are close to the receiver, whereas the third one is further away. Assume that the target rates can be achieved by the unique following power allocation: $p_1 = p_2= (P_{\max},P_{\max},0)$ and $p_3=(0,0,P_{\max})$. This happens for example if $R^t_3 = {1\over 3}f({P_{\max}g_{33}\over N_0})$, and $R^t_3> f({P_{\max}g_{33}\over N_0+P_{\max} g_{i3}})$, $i=1,2$ (in words, the third link cannot accomodate any kind of interference). Now the problem stems from the fact that if transmitters 1 and 2 select their allocation using PP first, then they would pick $p_1 =(P_{\max},P_{\max},0)$, $p_2 = (0,0,P_{\max})$. The third transmitter on the other hand needs to be {\it alone} in a slot to be satisfied (i.e., to achieve its target rate), but it cannot, and hence remains silent. The issue with IPP is actually common to all distributed power control protocols. A transmitter that causes low interference to others, but that is strongly interfered by others, has difficulties indicating its state to others through power control. Similarly, a link that cannot suffer much from interference has difficulties in gauging the impact of its power allocation on other links. 

The proposed solution to this problem marries Power Packing principle and randomization. PP is used to (quickly) reach a feasible power allocation, when PP can indeed go there. We believe that in most cases PP actually finds a feasible allocation. Randomization only helps IPP algorithm when the latter cannot converge to the desired allocation. Thus, the proposed schemes can be thought of as the {\it perturbed} version of IPP algorithm.

\subsection{Iterative Perturbed Binary PP algorithm}

The key idea of the algorithm is to force transmitters that are satisfied but whose power allocation is not compatible with any globally feasible allocation to explore other power allocations. This exploration is here triggered when unsatisfied transmitters create enough interference so that the target rate of satisfied links cannot be achieved anymore. The algorithm works as follows. First we generate a sequence of transmitters selected to update their power allocation over the frame. Then the power updates satisfy rules that we describe below.

{\it Update sequences.} As for IPP and IBPP algorithms, transmitters update their power allocation sequentially. The sequence of updates is driven by $s=(s[t])_{t\ge 0}$, assumed here to satisfy the following property:
\begin{itemize}
\item[(P2)] $(s[t])_{t\ge 0}$ is a stationary ergodic Markov chain with state space $\{1,\ldots,N\}$, such that $\mathbb{P}[s[t]=i]>0$ for all transmitter $i$.
\end{itemize}
A sequence satisfying (P2) may be generated when updates are triggered by independent Poisson clocks of identical rates, say $\nu$, at the various transmitters. To be more specific, when the clock of a transmitter ticks, the latter starts a power update at the next frame. When the common clock rate is relatively low (compared to the inverse of the frame duration), it is very unlikely that updates at two transmitters overlap (the convergence of our algorithms holds even in case of unfrequent update overlapping -- in fact convergence takes a finite number of updates, and so we just need that such sequence of updates occurs with positive probability). Under the above scenario, observe that for each new update, the selected transmitter is selected uniformly at random, so that (P2) is satisfied.  Note also that the time between updates occur at instants of a Poisson process of mean rate $N\nu$.

{\it Updating rules.} When an unsatisfied transmitter is picked for a possible update, it picks a power allocation as per the Binary-PP algorithm with probability (w.p.) $(1-\alpha_1)$, and picks a random allocation w.p. $\alpha_1$. A random power allocation can be obtained by using power $P_{\max}$ on each slot w.p.~1/2 independently of the power levels used in other slots. When a satisfied transmitter $i$ is selected, it checks whether its target rate has been achieved because of its own power allocation decision in the past ($\beta_i = 1$) or because of changes in the power allocation by other transmitters ($\beta_i = 0$). It can for example happen that an other transmitter decided to remain silent (applying Binary-PP algorithm), which made $i$ satisfied. In this case, the power allocation used by $i$ might not be compatible with any globally feasible allocation, and transmitter $i$ should explore other allocations. Thus in the algorithm, when $\beta_i=0$, $i$ does not update its power allocation w.p.~($1-\alpha_2)$, and chooses a random power allocation w.p. $\alpha_2$. Parameters $\alpha_1$ and $\alpha_2$ characterize the level of randomization in the algorithm. In what follows, we always assume that $0< \alpha_1, \alpha_2 <1$. When they are small, the algorithm is close to the initial IBPP algorithm, and converges very fast to a feasible allocation if IBPP can find one, but the algorithm would then take more time to identify a feasible allocation that IBPP cannot reach. The pseudo-code of the algorithm is presented below.  

\begin{separation}
\vspace{-0.07cm}
{\bf Iterative Perturbed Binary-PP (IPB-PP) algorithm.} 
%\vspace{-0.5cm}\separator
%\vspace{-0.2cm}
\begin{itemize}
\item[Input:] target rate vector $R^t$, update sequence $s$, power allocation $p[0]$, $\beta[0]\in\{0,1\}^K$.
\item[For each step $t\ge 1$:] Let $i=s[t]$.
\item[1.] Tx $i$ measures interference levels $I_{i}(p[t-1])$ in the different slots.
\item[2.] Tx $i$ updates its power allocation to $p_i[t]$:
\begin{itemize}
\item[(i)] If $R_i(p[t-1]) < R_{i}^t)$, then $p_i[t]$ is obtained as per BPP algorithm with inputs $R_{i}^t$ and $I_{i}(p[t-1])$ w.p. $1-\alpha_1$ and is a random power allocation w.p. $\alpha_1$; 
\item[(ii)] Else  If $(\beta_i[t-1]=0$), $p_i[t]=p_i[t-1]$ w.p. $(1-\alpha_2)$ and $p_i[t]$ is random w.p. $\alpha_2$;\\
${}\quad{\ }\quad{}$ Else $p_i[t]=p_i[t-1]$.
\end{itemize}
\item[4. ] Tx $i$ sets $\beta_i[t]=1$ if $R_i(p_i[t],p_{-i}[t-1])\ge R_i^t$, and $\beta_i[t]=0$ otherwise.
\end{itemize}
\end{separation}

We prove the convergence of IPB-PP under the following assumption. Let $U(p)$ denote the set of unsatisfied links under binary power allocation $p$.
\begin{itemize}
\item[(A1)] For any power allocation $p$ such that $U(p)\neq \emptyset$ 
and $U(p)\neq \{1,\ldots,N\}$, either there exists $i\in U(p)$ such that for\\ 
$R(P_{\max},p_{-i}) \ge R_i^t$, or for $p'$ such that $p_i'=P_{\max}{\bf 1}$ for all $i\in U(p)$ and $p_i'=p_i$ for $i\notin U(p)$, $U(p)\subsetneq U(p')$.
\end{itemize}
The assumption states that for any given power allocation $p$, either there exists a unilateral change in the power allocation of an unsatisfied transmitter that makes it satisfied, or when unsatisfied transmitters all select $P_{\max}{\bf 1}$, at least one other link becomes unsatisfied.

\begin{theorem} \label{thm:IBPP1} If there exists a binary power allocation $p^\*$ such that 
$R(p^\*)\ge R^t$, if (A1) holds, then from any initial condition, 
IPB-PP algorithm converges almost surely to a power allocation ${p}$ such that $R({p})\ge R^t$.
\end{theorem}

The previous theorem does not lead to the rate-optimality of IPB-PP algorithm. Although the algorithm performs well in practice, there are still some target rate vectors that it cannot reach. This is typically the case where one link has very low target rate, in which case, assumption (A1) may not be satisfied (the corresponding transmitter can be hardly affected by interference). Next we propose a rate-optimal algorithm whose principles are similar to those of IPB-PP algorithm.

\subsection{Interference-Triggered algorithm}

The next algorithm follows the same design principles as IPB-PP algorithm. However, the way satisfied transmitters are forced to explore other power allocations is different: they explore new allocations if they perceive significant changes in interference. More precisely, exploration is triggered when the change in the sum of the interference measured in the various slots exceeds a threshold $\delta$. The pseudo-code of this new algorithm is presented below.

\begin{separation}
\vspace{-0.07cm}
{\bf Interference-Triggered IPB-PP (IT-IPB-PP) algorithm.} 
%\vspace{-0.5cm}\separator
%\vspace{-0.2cm}
\begin{itemize}
\item[Input:] target rate vector $R^t$, update sequence $s$, power allocation $p[0]$, previous interference levels $I^{\rm last}[0]\in\mathbb{R}_+^K$.
\item[For each step $t\ge 1$:] Let $i=s[t]$.
\item[1.] Tx $i$ measures interference levels $I_{i}(p[t-1])$ in the different slots.
\item[2.] Tx $i$ updates its power allocation to $p_i[t]$:
\begin{itemize}
\item[(i)] If $R_i(p[t-1]) < R_{i}^t)$, then $p_i[t]$ is obtained as per BPP algorithm with inputs $R_{i}^t$ and $I_{i}(p[t-1])$ w.p. $1-\alpha_1$ and is a random power allocation w.p. $\alpha_1$; 
\item[(ii)] Else If $|\sum_{m=1}^M I_{im}(p[t-1]) - I^{\rm last}_{i}[t-1]| > \delta$, $p_i[t]=p_i[t-1]$ w.p. $(1-\alpha_2)$ and $p_i[t]$ is random w.p. $\alpha_2$;\\
${}\ \  \quad\quad{}$ Else $p_i[t]=p_i[t-1]$.
\end{itemize}
\item[4. ] Tx $i$ sets $I^{\rm last}_{i}[t] = \sum_{m=1}^M I_{im}(p[t-1])$.
\end{itemize}
\end{separation}

We prove the convergence of the algorithm under the following assumption on $\delta$.
\begin{itemize}
\item[(A2)] For every set $U \not= \emptyset$ or $\{1,\ldots,N\}$,
there exists a set $U'\not= \emptyset$ satisfying (1) $U \cap U' = \emptyset$ and
(2)~for every $j \in U'$, $M P_{\max}\sum_{i\in U}g_{ij} > \delta$.
\end{itemize}

Assumption (A2) states that any set $U$ of transmitters can be "heard" by at least one link in $U^c=\{1,\ldots,N\}\setminus U$. Note that as long as $g_{ij} > 0$ for every $i,j$, for any $\delta>0$, one can find a frame size $M$ (large enough) such that (A2) is satisfied. In this sense, the assumption is not restrictive: one may choose $\delta$ depending on the sensitivity of receivers, and then tune $M$ so that (A2) holds. 

\begin{theorem} \label{thm:ITIBPP} If there exists a binary power allocation $p^\*$ such that 
$R(p^\*)\ge R^t$, and if with our choices of $\delta$ and $M$, (A2) holds, then from any initial condition, 
IT-PIB-PP algorithm converges almost surely to a power allocation ${p}$ such that $R({p})\ge R^t$.
\end{theorem}

The above theorem states that if $R^t\in \bar{\cal S}_M$ and if (A2) holds, then IT-IPB-PP algorithm converges to a feasible power allocation. Now combining, this result with that of Lemma 3.1, we deduce that IT-PIB-PP is approximately rate-optimal. To be more precise, for $\epsilon\in (0,1)$, we say here that an algorithm is $\epsilon$-rate-optimal if it can achieve any rate vector $R^t$ such that $R^t+\epsilon {\bf{1}}\in {\cal R}^{\sched}$. 

\begin{corollary}
For any $\epsilon >0$, and any threshold $\delta>0$, there exsists a frame size $M(\epsilon,\delta)$ such that if $M\ge M(\epsilon,\delta)$, IT-IPB-PP algorithm is $\epsilon$-rate-optimal.
\end{corollary}

\section{Throughput-optimality}

In the previous section, we developed an approximately rate-optimal and fully distributed scheduling scheme. We now turn our attention to scenarios where each transmitter is equipped with an infinite buffer where it stores packets before sending them, and we address the design of throughput-optimal and distributed scheduling algorithms. We first describe our assumptions on the arrival processes, and on the notion of system stability.

{\it Arrival processes.} 
We assume that packets arrive in transmitter-$i$'s buffer according to an i.i.d. process.
Let $A_i[t]$ denote the number of bits arriving in transmitter-$i$'s 
buffer during frame $t$. $(A_i[t])_{t\ge 0}$ forms an sequence of i.i.d. 
random variables such that $A_i[t]\le A<\infty$ for all $i$ and $t$. 
The mean arrival rate (per frame) at transmitter $i$ is denoted by 
$\lambda_i = \mathbb{E}[A_i[t]]$. Let $\lambda=(\lambda_1,\ldots,\lambda_N)$. Finally, we assume that arrival processes 
are independent across transmitters.

{\it Stability.} Let $Q_i[t]$ denote the number of 
bits in transmitter-$i$'s buffer at the beginning of frame $t$. 
It evolves as: $Q_i[t+1]=\max(0,Q_i[t]+A_i[t]-S_i[t])$, where $S_i[t]$ 
is the number of bits sent during frame $t$.
Let $B$ denote the time required to empty all queues, i.e., $B = \inf\{u: u\ge 0, Q_i[u] = 0, \forall i\}$. We say that the system is stable if $\mathbb{E}[B|Q[0]] <\infty$ for all initial queue vector $Q[0]=(Q_1[0],\ldots,Q_N[0])$ such that $Q_i[0]<\infty$, for all $i$. We say that an algorithm is $\epsilon$-throughput optimal if it stabilizes the system whenever $\lambda+\epsilon {\bf{1}} \in {\cal R}^{\sched}$.

We use IT-IPB-PP algorithm to design approximately throughput-optimal and fully distributed scheduling schemes. 

\subsection{Known arrival rates} 

If each transmitter $i$ is aware of its arrival rate $\lambda_i$, this design is straightforward: each transmitter $i$ selects a target rate $R_i^t$ slightly bigger than $\lambda_i$, and we then run the IT-IPB-PP algorithm with these target rates, even when its queue is empty (using dummy packets). Under this strategy, after convergence of the IT-IPB-PP algorithm (which occurs after a finite time with finite mean), queues behaves independently and each of them has an arrival rates strictly less than its fixed service rate, which ensures stability. Next we make these statements precise.  

\begin{lemma}\label{lem:thru} Let $M$ be a frame size such that IT-IPB-PP is $(\epsilon/2)$-rate-optimal. Assume that $\lambda-\epsilon{\bf{1}} \in {\cal R}^{\sched}$. Then under IT-IPB-PP algorithm with target rate vector $R^t = \lambda + (\epsilon/2){\bf{1}}$, the system is stable.
\end{lemma}

The above lemma simply states that IT-IPB-PP algorithm provides an $\epsilon$-throughput-optimal algorithm, if each transmitter knows its arrival rate. $\epsilon$ can be made as small as desired by increasing the frame size $M$.

\subsection{Unknown arrival rates}
When the arrival rate $\lambda_i$ is not known, transmitter $i$ estimates it. When its estimate is precise enough, it selects a target rate appropriately (again slightly bigger than its estimated arrival rate) and then runs the IT-IPB-PP algorithm with this target rate. More precisely, for any $i$, let $\lambda_i[t]={1\over t}\sum_{s=1}^tA_i[s]$ and let $\mu = \epsilon/8$. Further define the interval $e_k = [2(k-1)\mu, 2k\mu)$. The target rate vector is continuously updated as follows: for any $i$,
$$
\hbox{if }\lambda_i[t] \in e_k,\hbox{ then }R_i^t[t] = (4k+1)\mu/2.
$$ 
When $\lambda_i$ lies in the interior for some $e_k$, since $\lambda_i[u]\to \lambda_i$ a.s. as $u\to\infty$, after a finite time $T_i$, $R_i^t[t]$ does not change anymore. In appendix we briefly explain how the case $\lambda_i = 2k\mu$ can be handled.  The following lemma then relies on the facts that $R_i^t[T_i]\in (\lambda_i,\lambda_i+\epsilon/2)$ and $\mathbb{E}[T_i]<\infty$ (proved in appendix):

\begin{lemma}\label{lem:thru2} Let $M$ be a frame size such that IT-IPB-PP is $(\epsilon/2)$-rate-optimal. Assume that $\lambda-\epsilon{\bf{1}} \in {\cal R}^{\sched}$, and that IT-IPB-PP algorithm is executed jointly with the above target rate update algorithm. Then  the system is stable.
\end{lemma} 

According to the above lemma, the proposed joint target rate update and scheduling algorithm is $\epsilon$-throughput-optimal. It is worth remarking that this algorithm proceeds in three phases: in the first phase, each transmitter aims at identifying a target rate that is just strictly greater than the arrival rate of bits in its buffer; in the second phase, IT-IPB-PP algorithm finds a power allocation compatible with the target rate vector; and finally, transmitters apply this power allocation, and queues empty. Also note that our algorithm is not designed so as to adapt to changing traffic conditions (i.e., changes in the arrival rates). A way to devise adaptive algorithms would be to let each transmitter continuousloy updates its target rate, depending on its observed queue length. To study such queue length based algorithm, one would need to understand the interaction between dynamics of the queues and of our IT-IPB-PP algorithm, which would require a significantly more involved analysis.

\section{Numerical experiments}

\begin{figure}[htb]
  \centering
  \includegraphics[width=0.65\columnwidth]{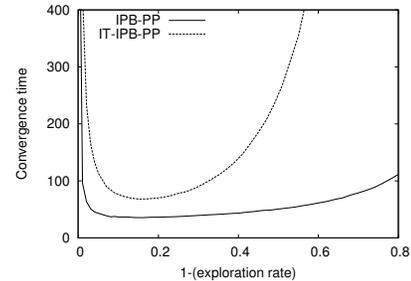}
\caption{Convergence time vs Exploration rate $\alpha_1$ -- $N=3=M$}
  \label{fig:1}
\end{figure}

In this section, we present simulation results to illustrate 
the rate-optimality of IPB-PP and IT-IPB-PP algorithms. For all experiments, the sensitivity parameter $\delta$ in IT-IPB-PP algorithm is fixed. We first experiment with a 3-link network. 
The network geometry is such that transmitters 
1 and 2 strongly interfere link 3, 
whereas transmitter 3 does not produce much interference, i.e., 
$g_{13}=g_{23}=60$, $g_{31}=g_{32}<1$, and the other gains are equal to 1. The target rate vector is chosen so that 
it cannot be reached by simple iterative Power Packing. It corresponds to a power allocation close to $p_1=p_2=(P_{\max},P_{\max},0)$, $p_3=(0,0,P_{\max})$. Fig. \ref{fig:1} 
shows the convergence time (in number of updates) of IPB-PP and IT-IPB-PP algorithms as 
a function of the exploration rate $\alpha_1$ (we choose $\alpha_2=\alpha_1$). The convergence time is averaged over 10,000 simulations starting from random power allocations. The convergence time rapidly grows either when the exploration rate is close to 0, or when it becomes too large. In the former, the algorithms behave like Binary-IPP, and cannot find a feasible allocation. In the latter, the algorithms get closer to a random search algorithm, and the convergence time explodes. Hence, in IPB-PP and IT-IPB-PP algorithms, it is clear that both Power Packing and randomization components are crucial: PP accelerates the convergence and randomization helps where PP fails at identifying a feasible allocation. It is worth noting that when the target rate vector can be achieved through simple Power Packing (without randomization), the convergence of the algorithm is very fast. 

\begin{figure}[t!]
\centering
\subfigure[Prop. of non-achieved rate vectors vs. frame size]{
\label{fig:M_frac}
\includegraphics*[width=0.48\columnwidth]{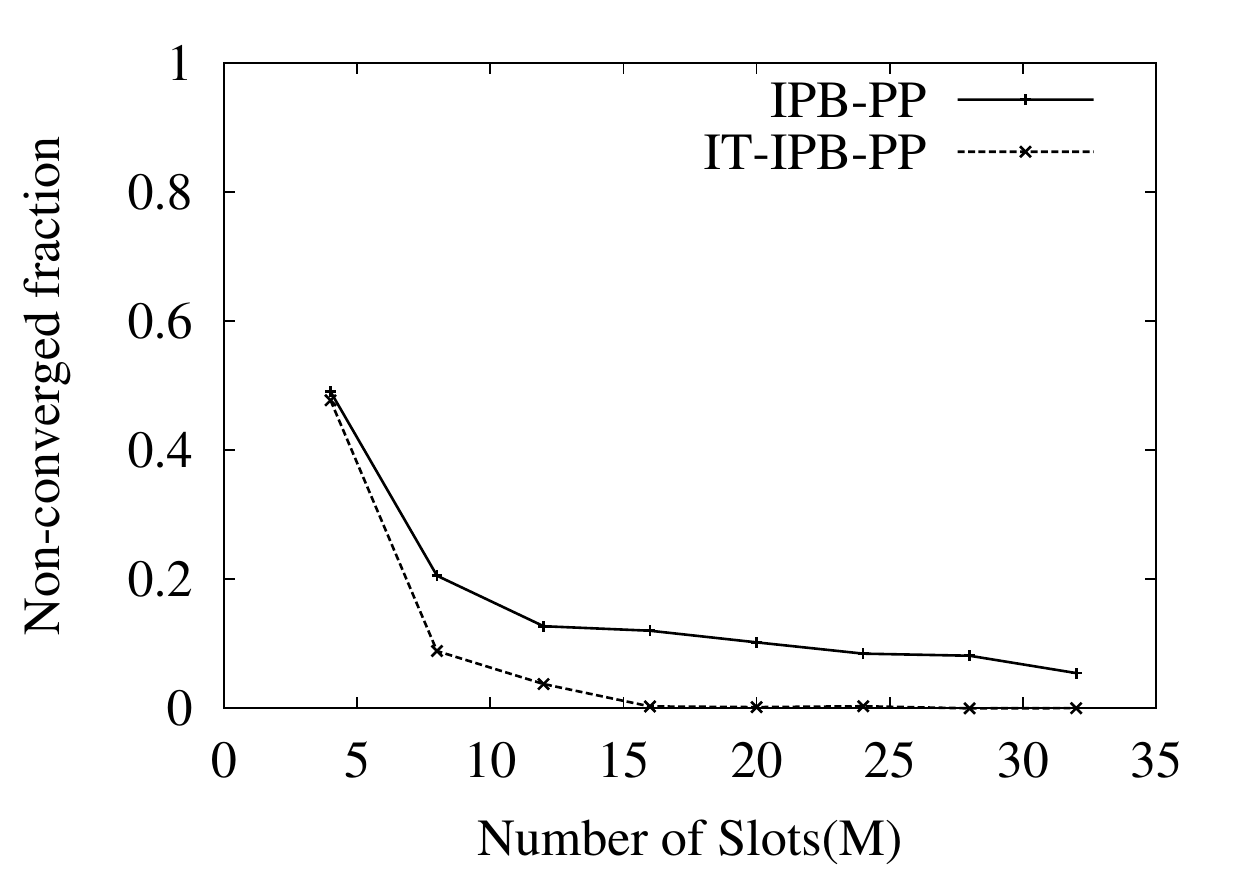}}
\subfigure[Avg. number of updates for convergence vs. frame size]{
\label{fig:M_steps}
\includegraphics*[width=0.48\columnwidth]{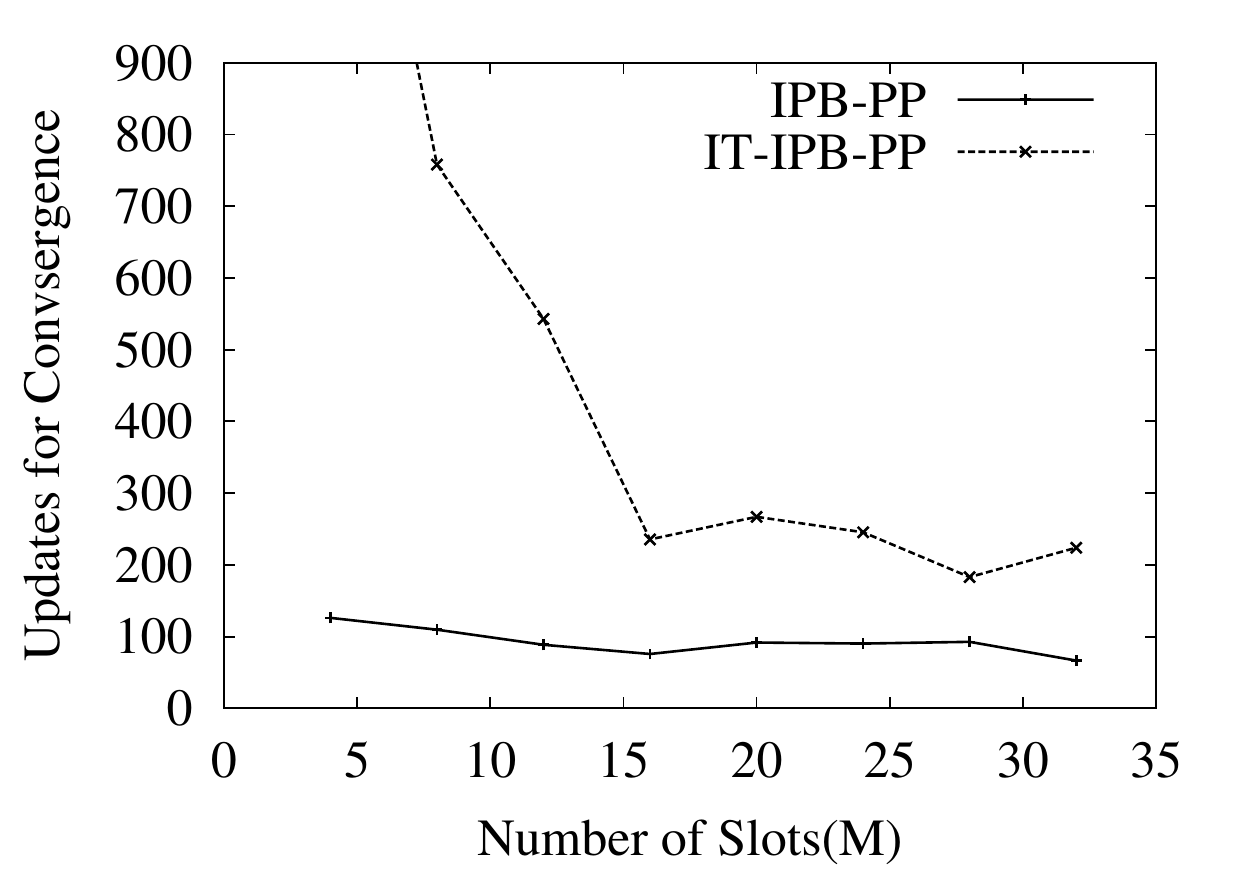}} \hspace{-3mm}
\caption{Performance of IPB-PP and IT-IPB-PP - $\alpha_1=0.1$}
\subfigure[Avg. number of updates for convergence vs. exploration rate]{
\label{fig:eps_steps}
\includegraphics*[width=0.49\columnwidth]{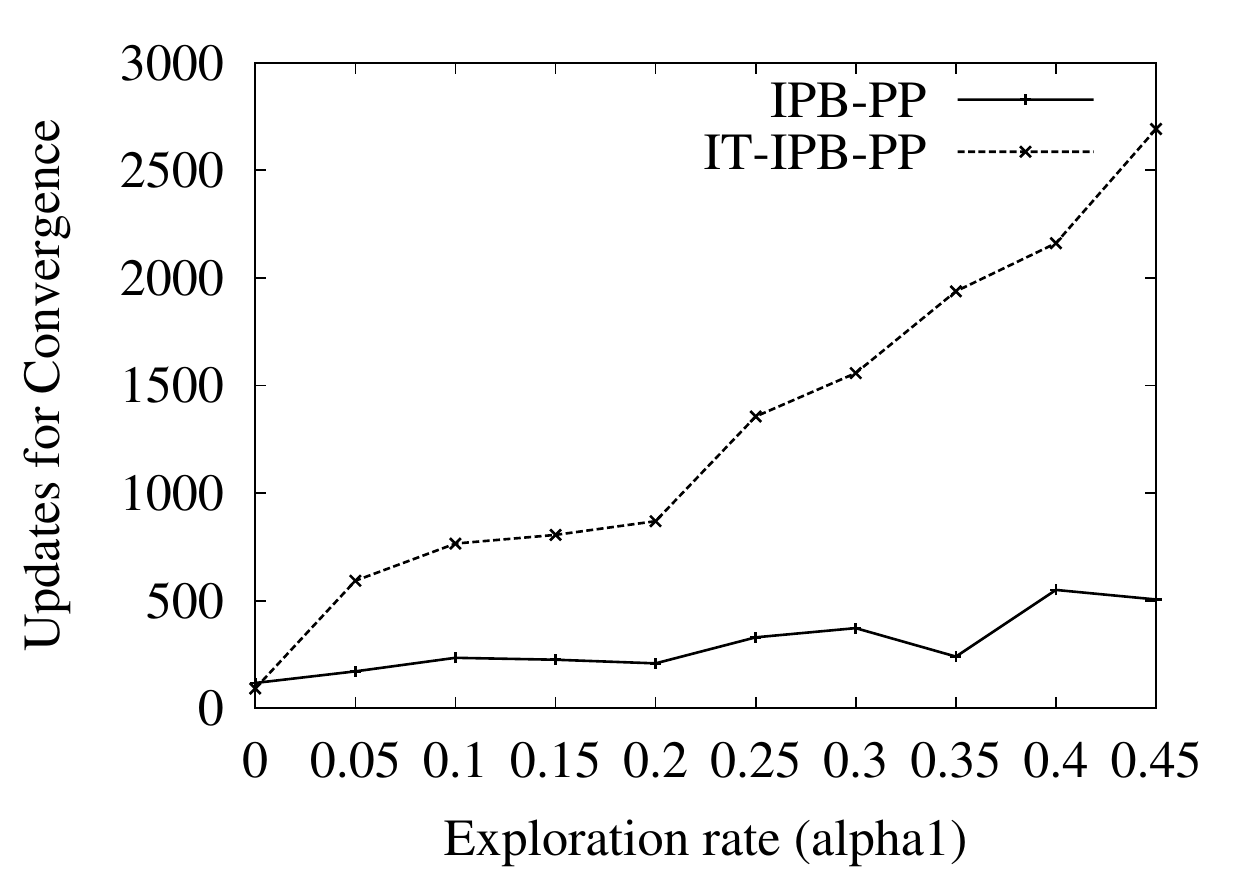}} \hspace{-3mm}
\subfigure[Prop. of non-achieved rate vectors vs. exploration rate]{
\label{fig:eps_frac}
\includegraphics*[width=0.49\columnwidth]{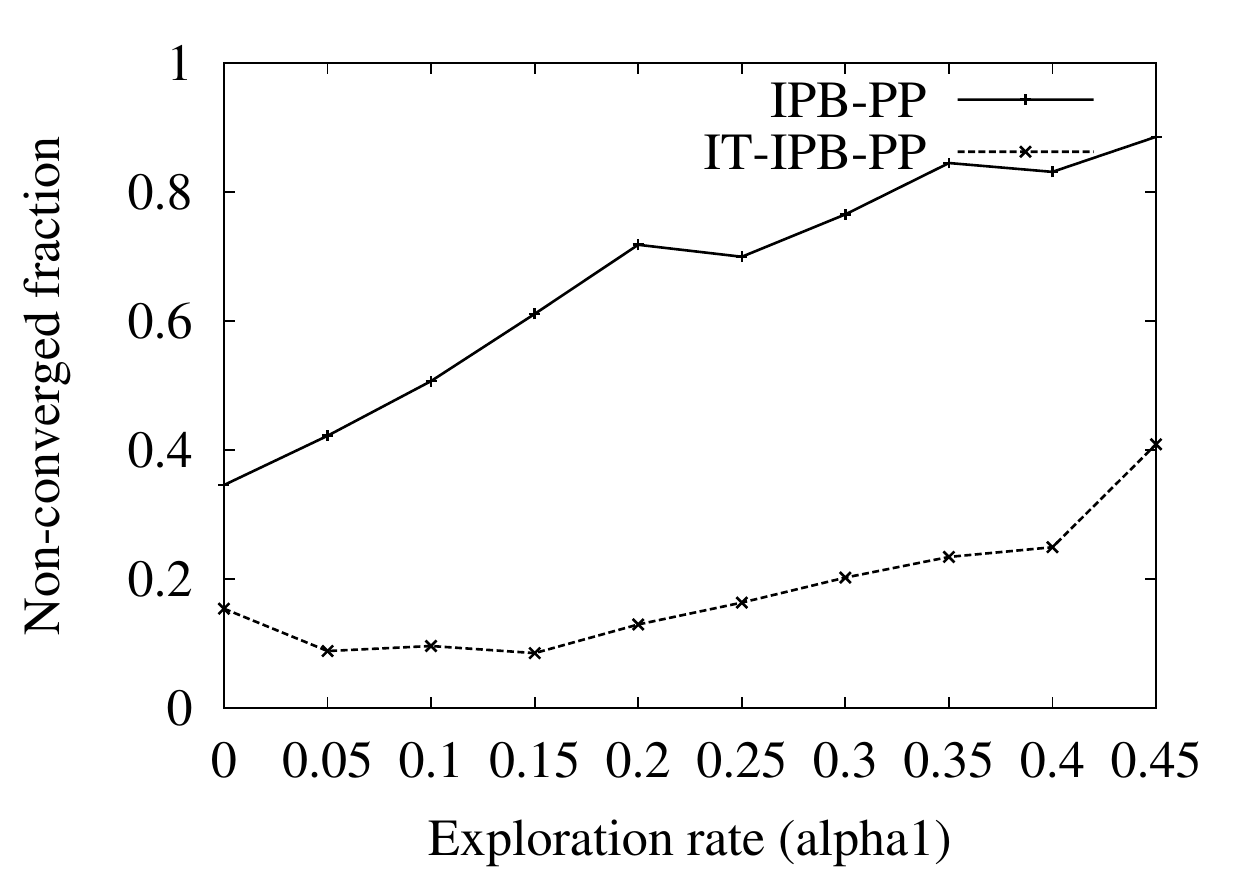}}
\caption{Performance of IPB-PP and IT-IPB-PP - $M=8$} 
\end{figure}

Next we consider randomly generated networks by placing 10 links on a 2D square (gains are computed using a path loss exponent equal to 3). For each generated network topology, we further generate $10^4$ target rate vectors in ${\cal S}_M$. For each vector we analyze the convergence time if the latter remains less than $10^4$ updates. We use two metrics for comparison: (1) the average (over topologies and rate vectors) number of updates required for convergence, given that it remains less than $10^4$, and (2) the proportion of rate vectors for which the algorithm does not converge in less $10^4$ updates. 

We first investigate the performance of our algorithms when the frame size $M$ varies. Here we fix $\alpha_1=\alpha_2=0.1$. Figure~\ref{fig:M_frac} shows that as $M$ increases, the proportion of rate vectors not reached by the algorithms rapidly decreases. For IT-IPB-PP, all vectors are achieved when $M=16$, illustrating the rate-optimality of the algorithm. Note that this is not the case of IPB-PP, as we predicted. Figure~\ref{fig:M_steps} shows how the convergence time varies with $M$. IT-IPB-PP seems to conevrge faster, and for both algorithms the convergence speed is increased when $M$ grows large. 

We now challenge our algorithms, and evaluate their performance when the frame size is not sufficient to guarantee the rate-optimality of IT-IPB-PP: we fix $M=8$, and vary the exploration rate $\alpha_1$. In Figure~\ref{fig:eps_steps}, we observe that in this case, the convergence time increases when $\alpha_1$ decreases, which again ilustrates the importance of the Power Packing component in the algorithms. In Figure~\ref{fig:eps_frac}, the proportion of rate vectors not reached by the algorithms within $10^4$ updates seem to increase as $\alpha_1$ increases, which indicates the negative effect of an aggressive random exploration.

\section{Conclusion}

This paper presents the first distributed scheduling algorithms that are optimal under the realistic SINR interference model, and do not require message passing among transmitters. The fact that algorithms combining such properties exist in surprising. Our solution is based on combining a simple power allocation strategy, and randomization techniques. Without randomization, the power allocation scheme could not, alone, achieve all parts of the throughput region (although numerical experiments show that it reaches a vast majority of it), and hence randomization is needed. We actually believe that randomization is always needed, i.e., no deterministic resource allocation scheme can be optimal. It would be interesting to formally establish this result. We are also interested in studying the convergence time of our iterative power allocation scheme, and its impact on actual queueing delays.

\bibliographystyle{IEEEtran}
\bibliography{IEEEabrv,refs}

\appendix

\section*{Proof of Lemma \ref{lem:rr}} We prove $\lim_{M\to\infty}{\cal R}_M^{\pc}={\cal R}$. $\lim_{M\to\infty}\bar{\cal S}_M={\cal R}^{\sched}$ can be proved analogously. Let $R\in {\cal R}$. Since ${\cal R}={\hbox{conv}}({\cal R}_1^{\pc})\subset \mathbb{R}^N$, by Caratheodory's theorem, there exist a finite set \\
$\{ r_1,\ldots,r^{N+1}\}$ of $(N+1)$ points in ${\cal R}_1^{\pc}$ and positive real number $\lambda_1,\ldots,\lambda_{N+1}$ such that: $R=\sum_{j=1}^{N+1}\lambda_jr_j$, and\\ $\sum_{j=1}^{N+1}\lambda_j=1$.

For any $j=1,\ldots,N+1$, let $p^j\in [0,P_{\max}]^{N}$ denote the vector representing power levels used by the various transmitters to achieve rate vector $r_j$. Now for $M\ge 1$, we propose the following power allocation across the $M$ slots of a frame: for all $j=1,\ldots,N+1$, power levels $p^j$ are used for ${\lfloor M\lambda_j\rfloor}$ slots (where $\lfloor x\rfloor$ is the largest integer smaller than or equal to $x$). The power allocation is arbitrary for the remaining slots. Using this power allocation, the achieved rate vector $U_M$ satisfies the following component-wise inequality: $U_M\ge \sum_{j=1}^{N+1} {\lfloor M\lambda_j\rfloor\over M}\times r_j$.

Note that $\lim_{M\to\infty}{\lfloor M\lambda_j\rfloor\over M}=\lambda_j$ for all $j$, and thus\\
$\lim_{M\to\infty}U_M\ge R$. \ep

\section*{Proof of Theorem \ref{th:ipp}}

We prove the convergence of IPP algorithm (the proof of the convergence of Binary-IPP is similar and easier). Let $s$ be a sequence of updates satisfying property (P1). Without loss of generality, we can assume that $s[2t]=1$ and $s[2t+1]=2$, for all $t\ge 0$ (i.e., transmitters alternatively apply PP algorithm). We denote by $p[t]$ the power allocation after the $t$-th update, and abuse the notation by writing $R[t]=R(p[t])$. It can be readily seen that (i) after both transmitters update once, i.e., for $t\ge 2$, the power allocation $p[t]$ is $\sigma$-repulsive (repulsive under the slot permutation $\sigma$), for a fixed given permutation $\sigma$ of slots; (ii) $R_1[2t]\in \{0,R_1^t \}$ and $R_2[2t-1]\in \{0,R_2^t \}$ for any $t\ge 1$. Observation (i) can be easily proved by induction on $t$. Given the permutation $\sigma$, we introduce the following notation: let $p_1,p_1'\in [0,P_{\max}]^M$, we write $p_1\le_\sigma p_1'$ if for all $m$, $p_{1\sigma(m)} \le p_{1\sigma(m)}'$.

Now let $R^t \in {\cal R}_M^{\rm{IPP}}$. Let $p^{\star}=(p_1^{\star},p_2^{\star}$) be a $\sigma$-repulsive power allocation such that $R(p^{\star})=R^t$. We establish the convergence of IPP algorithm to $R^t$ by investigating various possible initial conditions. 

\noindent
{\it Case 1: At time $2t$, $p_1[2t]=0$.} This means that after an update, link-1 transmitter actually chooses to remain silent. Without loss of generality, we assume that $t=0$. We show by induction property $(N_t)$, stating that the sequence of power allocations is monotonically increasing, and that the target rates are {\it alternatively} achieved on links 1 and 2:\\
Property $(N_t)$: $p_1[2t-2]\le_\sigma p_1[2t]\le_\sigma p_1^\star$, $R_1[2t]=R_1^t$,  $p_2[2t-3]\le_\sigma p_2[2t-1]\le_\sigma p_2^\star$, $R_2[2t-1]=R_2^t$ (with the convention that $p_2[-1]=0$).

Let us prove $(N_1)$. We have $p_1[0]=0$. Then at time 1, link-2 transmitter applies PP algorithm, and selects allocation $p_2[1]$ such that $R_2[1]=R_2^t$. Observe that link 2 has no interference, so that $p_2[1]\le_{\sigma}p_2^{\star}$. Then link-1 transmitter updates its power allocation. Since $p_2[1]\le_{\sigma}p_2^{\star}$, it can choose an allocation $p_1[2]$ such that $R_1[2]=R_1^t$ and $p_1[1]\le_{\sigma}p_1^{\star}$. Thus $(N_1)$ holds. Now assume that $(N_{t-1})$ holds, and let us prove $(N_t)$. At time $2t$, link-1 transmitter updates its power, and since $p_2[2t-1]\le_{\sigma}p_2^{\star}$, it can choose an allocation $p_1[2t]$ such that $R_1[2t]=R_1^t$ and $p_1[2t]\le_{\sigma}p_1^{\star}$. The same argument applied for link-2 transmitter allows to finish the proof of $(N_t)$.

Now $p[t]$ is monotonically increasing (w.r.t. $\le_\sigma$), and hence it converges. Remark that because of monotonicity, after a finite number of updates, the number of slots used by transmitter 1 or 2 is fixed. Hence after these numbers are fixed, the transmitters just update power on a single slot (always the same). The updates correspond to the synchornous version of Foschini-Miljanic algorithm, and hence converge to a feasible solution. In other words, if $\lim_{t\to\infty}p[t]=p'$, then $R(p')=R^t$.   

The case where at time $2t+1$, $p_2[2t+1]=0$, is similar to Case 1.
 
\noindent
{\it Case 2: At time $2t+1$, $R_2[2t+1]=R_2^t$ and $R_1[2t+1]\le R_1^t$.} Without loss of generality, asume that $t=0$. Link-1 transmitter updates its power at time 2. There are two cases: \\
(i) $R_1(P_{\max}{\bf 1},p_2[1])<R_1^t$, in which case, $p_1[2]=0$, and we return to Case 1; \\
(ii) $R_1(P_{\max}{\bf 1},p_2[1])\ge R_1^t$, in which case, $p_1[2]$ is such that $R_1[2]=R_1^t$ and $p_1[2]\ge_\sigma p_1[0]$ (because $R_1[1]< R_1[2]$). Now we have $R_2[2]\le R_2^t$ because interference increased for link 2. We can show using induction arguments just as those used in Case 1 that the power allocation is monotonically increasing until one transmitter saturates and becomes silent. In the latter case, we are back to Case 1. If transmitters never reset their power, we have convergence towards the target rates (using the same argument as in Case 1).

The case where at time $2t$, $R_1[2t]=R_1^t$ and $R_2[2t]\le R_2^t$, is similar to Case 1. 

\medskip
\noindent
{\it Case 3: At time $2t+1$, $R_2[2t+1]=R_2^t$ and $R_1[2t+1]> R_1^t$.} In this case again, we can show convergence using monotonicty arguments exactly as in previous cases. Note that in this case, power allocations are monotonically decreasing. The case where at time $2t$, $R_1[2t]=R_1^t$ and $R_2[2t]> R_2^t$ is of course similar. \ep

\section*{Proof of Theorem \ref{thm:IBPP1}} 

Since the update sequence satisfies (P2), \\
$x[t]=(s[t],p[t],\beta[t])_{t\ge 0}$ is an homogenous Markov chain with finite state space. Observe that a set $\{ (s,p,{\bf 1}), s\in \{1,\ldots,N\}\}$ constitutes a communication class of this Markov chain if $R(p)=R^t$ (in such states, all links are satisfied, and do not update their allocations anymore). To prove the theorem, we just need to show that from any initial state, at least one of these communication classes are accessible, i.e., we construct a finite sequence of state transitions occuring with positive probability and leading to one of the aforementioned communication classes. To construct such a path, we use the fact that from any state, all transmitters are picked for possible a power update with positive probability. We also use the fact that if tx $i$ is chosen for an update, and either its target rate is not satisfied or its $\beta_i$ is equal to 0, then tx $i$ can pick {\it any} power allocation with positive probability.

W.l.o.g. we may assume that $U(p[0])\neq\emptyset$ and that there is no tx $i\in U(p[0])$ that can update its allocation and become satisfied. Indeed if this is not the case, we pick this tx. With positive probaility, it updates its power allocation to $P_{\max}{\bf 1}$ and becomes satisfied. We repeat this procedure: pick an unsatisifed tx that can become satisfied, and let it use $P_{\max}{\bf 1}$. From allocation $p[0]$, power levels have been only increased, and so we end up at a state where there is no unsatisifed tx that can become satisified by unilateral power update.

Our constructed path consists of phases, indexed by $k=0,1,2,...$. At the beginning of phase $k$, the set of unsatisfied links whose transmitters is not using allocation $P_{max}{\bf 1}$ is denoted by $V_k$. Phase $k$ consists in letting links from $V_k$ select allocation $P_{max}{\bf 1}$ (this occurs with positive probability because these links are not statisfied). Note first that $V_0\neq\emptyset$, for by assumption (A1), when tx from $U(p[0])$ use $P_{\max}{\bf 1}$, one satisfied link becomes unsatisfied. Such an update requires that at least a tx from $U(p[0])$ is able to increase interference, and hence is not already using allocation  $P_{\max}{\bf 1}$. We prove similarly that at the beginning of phase $k\ge 1$, either $V_k\neq\emptyset$ or every transmitter uses $P_{\max}{\bf 1}$. Assume that $V_k=\emptyset$, which means that all unsatisfied transmitters use $P_{\max}{\bf 1}$. Hence unsatisfied transmitters cannot change their power allocation either to become satisfied or to disatisfy one link. From (A1), we deduce that all links are unsatisfied, and hence all use $P_{\max}{\bf 1}$. In summary after at most $N$ phases, all links are unsatisfied and use allocation $P_{\max}{\bf 1}$.

After all links have become unsatisfied, we add the following phase. We pick tx one after the other once. When tx $i$ is picked, either it is unsatisfied, or due to power updates of previous transmitters, it has become satisfied, but the value of its parameter $\beta_i$ is 0 (because it was not picked earlier in this phase). Hence when tx $i$ is picked, it will update its power allocation. With positive probability, it selects $p_i^{\star}$. After the last tx is picked, each tx is satisfied, but the $\beta_i$'s may not be all equal to 1. Finally we add a last phase: only the tx's $i$ such that $\beta_i=0$ are picked, and they again select power allocation $p_i^{\star}$. Thus we constructed a positive probability path from any state to a state where every tx is satisfied and will not update its power again. \ep

\section*{Proof of Theorem \ref{thm:ITIBPP}} 

The proof is similar to that of Theorem \ref{thm:IBPP1}.\\ $y[t]=(s[t],p[t],I^{\rm last}[t])_{t\ge 0}$
is an homogeneous Markov chain with finite state space. A set $\{(s,p,I), s \in \{1,\ldots,N\}\}$
constitutes a communication class of this Markov chain if $R(p) \ge R^t$ (in such states, all links are satisfied, and do
not update their allocations anymore). We show that these classes are accessible, and from any state, we build a positive probability path towards one of these classes.

Let $y[0]$ be any initial state of the Markov chain. As in the proof of Theorem \ref{thm:IBPP1}, w.l.o.g. we may assume that $U(p[0])\neq\emptyset$ and that there is no tx $i\in U(p[0])$ that can update its allocation and become satisfied. Let $U_0 = U(p[0])$. By (A2), there exists a maximal set $U_1\not= \emptyset$ such that $U_0 \cap U_1 = \emptyset$ and for every $j \in U'$, $M P_{\max}\sum_{i\in U_0}g_{ij} > \delta$. Set $U_1$ is maximal in a sense that no set $U' \supset U_1$ satisfies (A2) for the set $U_0$. Similarly, we recursively define $U_w$ as the maximal set satisfying (A2) for the set $\cup_{\ell=1}^w U_{\ell}$ if $\cup_{\ell=1}^w U_{\ell} \not= \{1,\ldots,N\}$. Let $W$: $\cup_{\ell=1}^W U_{\ell} = \{1,\ldots,N\}$. Note that such $W$ exists and is less than or equal to $N$. Also note that the sets $U_{\ell}$'s define a partition of $\{1,\ldots,N\}$.

Our constructed path consists of phases, indexed by $k=0,1,2,...$. We show that we can build these phases with positive probability such that:\\ 
(i) In each phase, all tx's are selected once; tx's from $U_0$ are selected first, then tx's from $U_1$, and so on.
(ii) In phase $k$, the tx's not in $U_{k+1}\cup\ldots\cup U_W$ do not update their power allocation.
(iii) In phase $2k$, each tx $i$ updating its allocation selects $p_i=0$. In phase $2k+1$, they select $P_{\max}{\bf 1}$.\\
(iv) there is a phase that ends with all tx having power allocation $0$.\\
If this construction is valid, then from the state where all tx remain silent, we conclude as in the proof of Thoerem \ref{thm:IBPP1}: we let each tx pick $p_i^{\star}$, and run two phases to align the variables $I_i^{\rm{last}}$.

We now justify (i)-(ii)-(iii)-(iv). (i)-(ii) are immediate ($s$ satisfies (P2), and a tx may always pick the same allocation as before with positive probability). Note that because of (i), in each phase, each tx $i$ updates its value of $I_i^{\rm{last}}$. In phase 0, all tx's in $U_0$ are unsatisfied, they update their power, and all choose 0 with positive probability. At the beginning of phase 1, tx's in $U_0$ are unsatisfied, and pick $P_{\max}{\bf 1}$; after that, from (A2), any tx $i$ in $U_1$ noticed the increased interference in its parameter $I^{\rm{last}}$, and hence update its power allocation with positive probability - it selects $P_{\max}{\bf 1}$. In phase 2, tx's in $U_0$ are still not satisfied because from their perspective, interference has increased compared to that perceived initially; they can then update their allocations again and this time select 0 power. This will be noticed by tx's in $U_1$, that again will update their allocations and select 0 power. In phase 3, tx's in $U_0$ and $U_1$ will select allocation $P_{\max}{\bf 1}$, which will be noticed by tx's $U_2$. The latter will then select allocation $P_{\max}{\bf 1}$. Repeating this argument, we justify (iii). (iv) is readily deduced from (iii). \ep

\section*{Proof of Lemma \ref{lem:thru}} 

Let $T$ be the time at which the IT-IPB-PP has converged. Since $T$ is the absorbing time of a finite state Markov chain, we have $\mathbb{E}[T]<\infty$. Now at $T$, a worst case (sample-path wise) is obtained by assuming that in each queue $i$, there are $AT+Q_i[0]$ bits to be served. From $T$, queues behave independently, and are also independent of the r.v. $T$. Thus the system is stable if and only if each queue is stable. It remains to prove that each queue $i$ is stable. W.l.o.g., assume that at time 0, queue $i$ has $AT+Q_i[0]$ bits to be served, and let $B_i=\inf\{u:Q_i(u)=0\}$. Define $\lambda_i[u]={1\over u}\sum_{s=1}^uA_i[s]$. Let $\delta=R_i^t-\lambda_i >0$. We have:
\begin{align*}
\mathbb{P}[B_i\ge u] & \le \mathbb{P}[ AT+Q_i[0]+ u\lambda_i[u] \ge u R_i^t]\\
&\le \mathbb{P}[ {AT+Q_i[0]\over u} + \lambda_i[u]-\lambda_i \ge \delta]\\
&\le \mathbb{P}[ {AT+Q_i[0]\over u} \ge {\delta\over 2}]+\mathbb{P}[ \lambda_i[u] -\lambda_i \ge {\delta\over 2}]\\
&\le \mathbb{P}[ {AT+Q_i[0]\over u} \ge {\delta\over 2}]+c_1e^{-c_2u},\\
\end{align*}
where the last inequality is obtained using Hoeffding's inequality  ($c_1,c_2 >0$). We deduce that $\mathbb{E}[B_i]=\sum_{u=1}^\infty \mathbb{P}[B_i\ge u]<\infty$, and queue $i$ is stable.\ep

\section*{Proof of Lemma \ref{lem:thru2}}

We just need to prove here that $\mathbb{E}[T_i]<\infty$ and $R_i^t[T_i] \in (\lambda_i, \lambda_i+\epsilon/2)$. After establishing these results, we can apply the same proof as that of Lemma \ref{lem:thru}. Indeed, note that after $\max_iT_i$, at each transmitter, the target rate is fixed and greater than the arrival rate; also observe that $\mathbb{E}[ \max_iT_i] \le \sum_i \mathbb{E}[T_i]$.

We only consider that $\lambda_i$ lies in the interior of $e_k$ for some $k$.\footnote{The case where $\lambda_i$ may lie on the boundary of some $e_k$
can be handled similarly by choosing a slightly more complex target rate update algorithm: we consider two partitions of $\mathbb{R}^+$, $(e_k)_{k\ge 1}$ and $(f_k)_{k\ge 0}$ where $f_k=e_k+\mu/2$ for $k\ge 1$ and $f_0=[0,\mu/2)$. We consider the same rate update, but switch partition when $\lambda_i[t]$ falls into a different interval than that of $\lambda_i[t-1]$. Using this, after $\lambda_i[t]$ concentrates around $\lambda_i$, we do not switch partition anymore, and $\lambda_i$ lies in the interior of an interval of the partition.} 
Thus, there exists $\delta > 0$ such that $\delta$-neighborhood of $\lambda_i$ lies in $e_k$. Let $T_{\delta} = \inf\{t: \sup_{u\ge t}|\lambda_i[u]-\lambda_i| < \delta\}$.
Note that for every $t \ge T_\delta$, $\lambda_i[t] \in e_k$ and thus $R_i^t[t] = (4k+1)\mu/2$. Also observe that $T_i\le T_{\delta}$. We show that $\mathbb{E}[T_\delta] < \infty$.
Consider $\mathbb{P}\{T_\delta > t\}$ and note that
\begin{align*}
 \mathbb{P}\{T_\delta > t\} &= \mathbb{P}\{\cup_{u = t+1}^\infty\{|\lambda_i[u]-\lambda_i| \ge \delta\}\}\\
 &\le \sum_{u=t+1}^\infty \mathbb{P}\{|\lambda_i[u]-\lambda_i| \ge \delta\} \\
 &\le \sum_{u=t+1}^\infty c_1 e^{-c_2 u} \le \frac{c_1}{c_2}e^{-c_2 t}.
\end{align*}
The last inequality follows from Hoeffding's inequality ($c_1,c_2 >0$).
Now, $E[T_\delta] = \sum_{t=1}^\infty \mathbb{P}\{T_\delta \ge t\} \le \sum_{t=1}^\infty \frac{c_1}{c_2}e^{-c_2 t} < \infty$. Hence, $\mathbb{E}[T_i]<\infty$. Finally, from the fact that $\lambda_i\in e_k$ and $R_i^t[T_i]=(4k+1)\mu/2$, we simply deduce that $R_i^t[T_i] \in (\lambda_i, \lambda_i+\epsilon/2)$.

\ep

\end{document}